\documentclass[12pt]{article}

\usepackage{amsthm,amssymb,amsmath}
\usepackage[american,british]{babel}

\newtheorem{Th}{Theorem}
\newtheorem{pro}{Proposition}

\renewcommand{\a}{{\operatorname{affine}}}

\newcommand{\dif}{\operatorname{d}}
\renewcommand{\t}{{\operatorname{t}}}
\newcommand{\I}{\operatorname{i}}

 \newcommand{\res}{\operatorname{Res}}
 \newcommand{\norm}[1]{\left\Vert#1\right\Vert}

\newcommand{\diag}{\operatorname{diag}}

\newcommand{\bt}{{\boldsymbol t}}
\newcommand{\bx}{{\boldsymbol x}}
\newcommand{\be}{{\boldsymbol e}}

\newcommand{\bu}{{\boldsymbol u}}

\newcommand{\bpsi}{{\boldsymbol \psi}}
\newcommand{\bG}{{\boldsymbol g}}
\newcommand{\C}{{\mathbb C}}
\newcommand{\V}{{\mathbb V}}
\newcommand{\R}{{\mathbb R}}
\newcommand{\Z}{{\mathbb Z}}

\newcommand{\D}{{\partial}}

\begin{document}
\title{From Ramond Fermions
\\ to Lam\'{e} Equations for
\\Orthogonal Curvilinear Coordinates}
\author{Manuel Ma\~nas$^{1,2}$ and
Luis Mart\'\i nez Alonso$^2$ \\ \\
 $^1$Departamento de Matem\'{a}tica
Aplicada y Estad\'{\i}stica\\ E. U. I. T. Aerona\'{u}tica, Universidad
Polit\'{e}cnica de Madrid\\ E28040-Madrid, Spain\\
 $^2$Departamento de F\'\i sica
Te\'orica II, Universidad Complutense\\ E28040-Madrid, Spain}


\date{{\bf email:} manuel@dromos.fis.ucm.es, luis@ciruelo.fis.ucm.es}


\maketitle

\begin{abstract}
We show how Ramond free neutral Fermi fields lead to a
$\tau$-function theory of BKP type which describes iso-orthogonal
deformations of  systems of ortogonal curvilinear coordinates. We
also provide a vertex operator representation for the classical
Ribaucour transformation.
\end{abstract}
\newpage%

\section{Introduction}

In \cite{BKP} the multicomponent BKP hierarchies of integrable
systems were introduced by using free neutral Ramond fermions, the
building blocks of the  {\em fermionic quark model} for current
algebras \cite{olive}. Some correlations functions of the theory,
the so called $\tau$-functions, satisfy a bilinear identity,
defined in terms of  vertex operators, which is equivalent to the
BKP hierarchy of integrable nonlinear partial differential
equations.

In this letter we show that this approach can be applied to the
theory of systems of orthogonal curvilinear coordinates, a
classical subject in differential geometry
\cite{Darboux1,Eisenhart1,Bianchi}, and its  iso-orthogonal
deformations. Let us remind the reader that since the last century
distinguished geometers, among others Gauss, Dupin, Binet, Lam\'{e},
Darboux, Egorov and Bianchi established the basis of this theory.
 In particular,
Darboux's book \cite{darboux} is a standard reference for the Lam\'{e}
equations \cite{lame}, which describe these systems, see also
\cite{Bianchi} for the problem of triply orthogonal systems of
surfaces. Geometers were also interested in those transformations
(symmetries) \cite{darboux,eisenhart} preserving the orthogonal
character of systems of curvilinear coordinates and, in particular,
a  fairly general transformation of this type was derived by
Ribaucour \cite{ribaucour}. Very recently the Lam\'{e} equations have
been integrated by means of the inverse scattering technique
\cite{zakharov} and algebro-geometric solutions in terms of theta
functions  have been constructed in \cite{krichever}. Moreover, it
has been found that the Lam\'{e} equations are relevant in the
classification problem of 2D topological field theories. In
particular, it was shown by Dubrovin \cite{dubrovin} that the
descomposable associativity equations of
Witten-Dijkgraff-Verlinde-Verlinde \cite{w,dvv} can be described in
terms of some particular subclass of Lam\'{e} systems, those of Egorov
type, a particular class of flat diagonal metrics
\cite{Darboux3,Egorov}. Further, in \cite{krichever}, the partition
function of the theory was found in terms of theta functions.

In this letter we give:
\begin{enumerate}
\item A $\tau$-function and bilinear equation representation of
the Lam\'{e} equations.
\item A correspondence between the vertex operator of the theory
and the classical Ribaucour transformation.
\end{enumerate}
Our method  is based on the Grassmannian formalism \cite{sw} as
well as on some identities for Baker functions coming from  Fay
identities for $\tau$-functions. For analogous results regarding
the $N$-component KP hierarchy, charged fermions, conjugate nets
and its transformations see \cite{dmmms}.

The layout of the paper is as follows. Next, in \S 2, we present a
brief account of the Lam\'{e} equations and the Ribaucour
transformation. Then, in \S 3 we remind the reader the quantum
field theory of the BKP hierarchy in terms of $\tau$-functions. In
\S 4 we introduce the Baker matrix and show the relation between
the BKP hierarchy and orthogonal curvilinear coordinates. Finally,
in \S 5 we identify the action of the vertex operator with the
classical Ribaucour transformation

\section{ Lam\'{e} equations for Orthogonal nets}

Systems of curvilinear coordinates $\bu:=(u_1,\dots,u_N)$ in the
Euclidean space $\R^N$ are determined by diffeomorphisms
$\bu\mapsto\bx(\bu)$, with $\bx=(x_1,\dots,x_N)$ being the
Cartesian coordinates in $\R^N$.

Curvilinear coordinates such that the coordinates lines are
orthogonal satisfy:
\[
\dfrac{\D\bx}{\D u_i}\cdot\dfrac{\D\bx}{\D u_j}=0,\quad i\neq j,
\]
and the  corresponding normalized  tangent vectors are given by
\[
\bG_i=\frac{1}{H_i}\dfrac{\D\bx}{\D u_i},\quad i=1,\dots,N.
\]
where
\[
H_i:=\norm{\dfrac{\D\bx}{\D u_i}},\quad i=1,\dots,N,
\]
are the so called Lam\'{e} coefficients. It turns out that  the
orthonormal frame $\{\bG_j\}_{j=1}^N$ satisfies
\begin{gather}
\dfrac{\D\bG_i}{\D u_j}-\beta_{ij}\bG_j=0,\quad i,j=1,\dots,
N,\; i\neq j,\label{g1}\\
\frac{\D\bG_i}{\D u_i}+\sum_{\substack{k=1,\dots,N\\
k\neq i}}
 \beta_{ki}\bG_k=0, \quad i=1,\dots,N.\label{g2},
\end{gather}
where $\beta_{ij}$ are the  rotation coefficients:
\[
\beta_{ij}:=\frac{1}{H_i}\dfrac{\D H_j}{\D u_i}.
\]

The compatibility of this system implies the so called Lam\'{e}
equations for the rotation coefficients:
\begin{gather}
\frac{\partial\beta_{ij}}{\D u_k}-\beta_{ik}\beta_{kj}=0,\;\;
i,j,k=1,\dotsc, N,\;
\text{with $i,j,k$ different},\label{lame1}
\\
\frac{\partial\beta_{ij}}{\D u_i}+
\frac{\partial\beta_{ji}}{\D u_j}+
\sum_{\substack{k=1,\dotsc,N\\ k\neq i,j}}
\beta_{ki}\beta_{kj}=0,\quad i,j=1,\dotsc,N,\;i\neq j.\label{lame2}
\end{gather}
The Cartesian coordinates  $(x_1,\dots,x_N)$ are recovered from the
following Laplace equations
\begin{gather*}
\frac{\D^2\bx}{\D u_i\D u_j}=\frac{\D \ln H_i}{\D u_j}\frac{\D \bx}{\D u_i}
+\frac{\D \ln H_j}{\D u_i}\frac{\D \bx}{\D u_j},
\quad i,j=1,\dotsc,N,\;\; i\neq j,\\
\frac{\D^2\bx}{\D u_i^2}+\frac{1}{2}\frac{\D (H_i^2)}{\D u_i}\sum_{k=1}^N
\frac{1}{H_k^2}\frac{\D\bx}{\D u_k}=0, \quad i=1,\dots,N,
\end{gather*}
once the Lam\'{e} coefficients are determined.

Given a system of orthogonal curvilinear coordinates it is of
interest to derive transformations providing a new set of
orthogonal curvilinear coordinates. A transformation of this type
was found in the last century and is known as the Ribaucour
transformation. It requires the introduction of a potential in the
following manner: given functions $\zeta_i$ such that
\begin{equation}\label{ji}
\frac{\D\zeta_i}{\D u_j}=\beta_{ij}\zeta_j,\quad i,j=1,\dots,N,\;
i\neq j,
\end{equation}
 one can define a potential $\Omega(\zeta,H)$ through the compatible
equations
\begin{equation}\label{potential}
\frac{\D\Omega(\zeta,H)}{\D u_i}=
\zeta_i H_i
\end{equation}

\newtheorem*{VF}{\textit{Ribaucour Transformation}}

\begin{VF}
Given solutions $\zeta_i$ of \eqref{ji}, $i=1,\dotsc,N$,
 new rotation coefficients ${\cal R}(\beta_{ij})$,
 orthonormal tangent vectors
${\cal R}(\bG_i)$, Lam\'e coefficients ${\cal R}( H_i)$ and flat
coordinates ${\cal R}(\bx)$
 are given by
\begin{equation}\label{ribaucour}
\begin{aligned}
{\cal R}(\beta_{ij})&=\beta_{ij}-
2\dfrac{\zeta_i}{\sum_{k=1}^N\zeta_k^2}\Big[\frac{\D\zeta_j}{\D
u_j}+
\sum_{\substack{k=1,\dots,N\\k\neq j}}\beta_{kj}\zeta_k\Big],\\
{\cal R}(\bG_i)&=\bG_i-2\dfrac{\zeta_i}{\sum_{k=1}^N\zeta_k^2}
\sum_{k=1}^N\zeta_k\bG_k.\\
{\cal
R}(H_i)&=H_i-2\dfrac{1}{\sum_{k=1}^N\zeta_k^2}\Big[\frac{\D\zeta_j}{\D
u_j}+
\sum_{\substack{k=1,\dots,N\\k\neq j}}\beta_{kj}\zeta_k\Big]\Omega(\zeta,H),\\
{\cal R}(\bx)&=\bx- 2\frac{1}{\sum_{k=1}^N\zeta_k^2}\Omega(\zeta,H)
\sum_{k=1}^N\zeta_k\bG_k.
\end{aligned}
\end{equation}
\end{VF}

\section{Ramond fermions and BKP hierarchies}

 The $N$-component BKP hierarchy can be
defined in terms of Ramond neutral free Fermi fields   as follows
\cite{BKP}. First, we introduce a set of anticommuting quantum
fields $\Phi_i(z)$, $i=1,\dots,N$, satisfying
\[
\{\Phi_i(z),\Phi_j(z')\}=\delta_{ij}\delta(z+z'),
\]
where $\delta(z)$ denotes the Dirac distribution on the unit circle
$S^1=\{z\in\C:|z|=1\}$. These quantum fields have a Laurent
expansion in $z$:
\[
\Phi_i(z)=\sum_{n\in\Z} z^n\Phi_{i,n},
\]
with
\[
\{\Phi_{i,n},\Phi_{j,m}\}=\delta_{i,j}\delta_{n,-m},\quad
i,j=1,\dots,N,\; n,m\in\Z.
\]
 The vacuum $|0\rangle$ and antivacuum $\langle 0|$ are defined
by the relations:
\begin{gather*}
\Phi_{i,n}|0\rangle=0,\quad n<0,\\
\langle 0|\Phi_{i,n}=0,\quad n>0,\\
\langle 0|\Phi_{i,0}\Phi_{j,0}|0\rangle=0
\end{gather*}

Now, we consider an infinite number of time labels:
\[
\bt=(\bt_1,\dots,\bt_N)\in\C^{N\raisebox{.4mm}{
\scriptsize$\cdot\infty$}},\quad
\bt_i:=(t_{i,1},t_{i,3},t_{i,5},\dotsc)\in\C^\infty,
\]
in terms of which we construct the operator:
\[
H(\bt)=:\frac{1}{2}\sum_{\substack{i=1,\dots,N\\n\in\Z\\l\geq 0}}
(-1)^{n+1}t_{j,2l+1}\Phi_{j,n}\Phi_{j,n-2l-1}.
\]
The quadratic products $\Phi_{i,n}\Phi_{j,m}$ when exponentiated
generate a Lie group $G$. Given a element $g\in G$, there is a set
of associated $\tau$-functions given by the following expectation
values
\begin{gather*}
\tau(\bt):=\langle 0|\exp(H(\bt))g|0\rangle,\\
\tau_{ij}(\bt):=\langle 0|\exp(H(\bt))g\Phi_{i,0}\Phi_{j,0}|0\rangle,\quad
i,j=1,\dots,N.
\end{gather*}
These correlations satisfy
\begin{gather*}
\tau_{ij}+\tau_{ji}=0,\quad i,j=1,\dots,N,\; i\neq j,\\
2\tau_{ii}=\tau,\quad i=1,\dots, N.
\end{gather*}
and the next bilinear equation ---which defines the $N$-component
BKP hierarchy---
\begin{equation}\label{bilinealBKP}
\sum_{k=1}^N\int_{S^1}\dfrac{\dif z}{2\pi \I z}
[X_k(z)\tau_{ik}(\bt)][X_k(-z)\tau_{jk}(\bt')]=
\sum_{k=1}^N\tau_{ki}(\bt)\tau_{kj}(\bt'), \; i,j=1,\dots,N,
\end{equation}
for suitable $\bt$ and $\bt'$. Here  we are using the following
vertex operators
\begin{gather*}
X_i(z):=\exp(\xi(z,\bt_i))\V_i(-z),\\
\xi(z,\bt_i):=\sum_{n=1}^\infty z^{2n+1} t_{i,2n+1},\quad
\V_i(z)f(\bt):=f(\bt+[1/z]\be_i),\\
\left[ 1/z \right]:= 2\left( \frac{1}{z}, \frac{1}{3z^3},
\frac{1}{5z^5}
\ldots \right),
\end{gather*}
 with  $\{\be_i\}_{i=1}^N$  being the canonical generators of
 $\C^N$, so that the
bilinear equation \eqref{bilinealBKP}  can be written as
\begin{multline}\label{bilinealBKP2}
\sum_{k=1}^N\int_{S^1}\dfrac{\dif z}{2\pi \I z}
\exp(\xi(z,\bt_k)-\xi(z,\bt'_k))\tau_{ik}(\bt-[1/z]\be_k)
\tau_{jk}(\bt'+[1/z]\be_k])]\\=
\sum_{k=1}^N\tau_{ki}(\bt)\tau_{kj}(\bt'),\; i,j=1,\dots,N.
\end{multline}

\section{From  BKP hierarchy to  Lam\'{e} equations}
The above $\tau$-function formulation of the bilinear equation of
the $N$-component BKP hierarchy allows for a useful Baker-function
description. For this aim we introduce a non-normalized wave
function $\varphi(z,\bt)$
\[
\varphi_{ij}(z,\bt):=\tau_{ij}(\bt-[1/z]\be_j)\exp(\xi(z,\bt_j)).
\]
The bilinear equation \eqref{bilinealBKP} becomes
\[
\int_{S^1}\frac{\dif z}{2\pi\I
z}\varphi(z,\bt)\varphi(-z,\bt')={\cal T}^\t(\bt){\cal T}(\bt'),
\]
with
\[
{\cal T}(\bt):=(\tau_{ij}(\bt)).
\]
Observe that
\[
\varphi(z,\bt)=\rho(z,\bt)\psi_0(z,\bt),
\]
where
$\psi_0(z,\bt):=\diag(\exp(\xi(z,\bt_1)),\dots,\exp(\xi(z,\bt_N)))$,
and $\rho$ has  the following asymptotic expansion
\[
\rho(z)\sim {\cal T}+\alpha z^{-1}+{\cal O}(z^{-2}),\quad z\to\infty.
\]
In order to get a Baker function one normalize the above wave
function by its dominant behavior at $z=\infty$; i. e., the Baker
function is
\[
\psi(z,\bt):={\cal T}^{-1}(\bt)\varphi(z,\bt),
\]
and the bilinear equation \eqref{bilinealBKP} becomes
\begin{equation}\label{bilbaker}
\int_{S^1}\frac{\dif z}{2\pi\I
z}\psi(z,\bt)\psi^\t(-z,\bt')=G(\bt)G^\t(\bt'),\quad G(\bt):={\cal
T}^{-1}(\bt){\cal T}^\t(\bt).
\end{equation}
The asymptotic behavior of the Baker function is, by construction,
\[
\psi(z,\bt)=\chi(z,\bt)\psi_0(z,\bt),
\]
where
\[
\chi(z,\bt)\sim 1+\beta(\bt) z^{-1}+{\cal O}(z^{-2}),\quad z\to\infty.
\]

A direct consequence of \eqref{bilbaker}, when $\bt=\bt'$, is that
the matrix $G(\bt)$ is orthogonal. Moreover, assuming that
  the Baker function is meromorphic inside the unit
circle with poles at $\{z_1,\dots,z_n\}$, we see that the integrand
in
\eqref{bilbaker} has poles at
$\{0\}\cup\{z_1,\dots,z_n\}\cup\{-z_1,\dots,-z_n\}$, and,
therefore, by evaluating the corresponding residues, we get:
\begin{multline*}
\psi(0,\bt)\psi^\t(0,\bt')+\sum_{i=1}^n\frac{1}{z_i}
\Big[
\res_{z_i}(\psi(z,\bt))\psi^\t(-z_i,\bt')-\psi(-z_i,\bt)\res_{z_i}(\psi^\t(z,\bt'))
\Big]\\=G(\bt)G^\t(\bt').
\end{multline*}
Transposing this relation we conclude that
\[
\sum_{i=1}^n\frac{1}{z_i}
\Big[
\res_{z_i}(\psi(z,\bt))\psi^\t(-z_i,\bt')-\psi(-z_i,\bt)\res_{z_i}(\psi^\t(z,\bt'))
\Big]=0;
\]
hence, it follows that
\[
\psi(0,\bt)\psi^\t(0,\bt')=G(\bt)G^\t(\bt').
\]
Therefore
\[
G^{-1}(\bt)\psi(0,\bt)= G^\t(\bt')(\psi^\t)^{-1}(0,\bt'),
\]
is a constant orthogonal matrix. We notice that \eqref{bilbaker}
determines $G$ up to $G\to GC$, $C\in\text{O}_N$, so that we can
set:
\[
G(\bt)=\psi(0,\bt).
\]

In order to derive linear systems for the Baker functions we will
use a Grassmannian like approach \cite{sw}.
 Observe that in terms
of the matrix function
\[
\Phi(z,\bt):=G^{-1}(\bt)\psi(z,\bt)
\]
the bilinear relation reads
\[
\int_{S^1}\frac{\dif z}{2\pi\I
z}\Phi(z,\bt)\Phi^\t(-z,\bt')=1.
\]
Consider now the affine space $W_\a$ of functions $w(z)$ such that
\[
\int_{S^1}\frac{\dif z}{2\pi\I
z}w(z)\Phi^\t(-z,\bt')=1,
\]
for all suitable $\bt^\prime$, and the linear space $W$ of
functions $w(z)$ such that
 \[
\int_{S^1}\frac{\dif z}{2\pi\I
z}w(z)\Phi^\t(-z,\bt')=0,
\]
for all suitable $\bt^\prime$.

We introduce  time evolutions of these spaces  by $W_\a\mapsto
W_\a(\bt)=W_\a\psi_0(z,\bt)$ and $W\mapsto W(\bt)=W\psi_0(z,\bt)$.

The main properties of these  spaces are
\begin{pro}\label{asymptotic}
\begin{enumerate}
  \item The only element in $W_\a(\bt)$ with asymptotic expansion
  of the form
  \[
  M(\bt)+{\cal O}(z^{-1}),\quad \text{when } z\to\infty
  \]
is $\Phi(z,\bt)\psi_0(z,\bt)$.

  \item The linear space $W(\bt)$ has no elements with asymptotic
  expansion
  \[
  R(\bt)z^{-n}+{\cal O}(z^{-n-1}),\quad R\neq 0,\; n\geq 0,
  \]
when $z\to \infty$.
\end{enumerate}
\end{pro}
With these at hand we can show that the BKP hierarchy constitutes a
set of iso-orthogonal deformations of orthogonal nets. We shall
denote by $\bpsi_i$  and $\bG_i$ the $i$-th row of the matrices
$\psi$ and $G$, and use the notation $u_i=t_{i,1}$ as well as
$E_{kl}$ for the matrix $(\delta_{ik}\delta_{lj})$, and
$P_i=E_{ii}$.
\begin{Th}
\begin{enumerate}
\item The vectors $\bpsi_i(z,\bt)$  satisfy:
\[
\dfrac{\D\bpsi_i}{\D u_j}-\beta_{ij}\bpsi_j=0,\quad i,j=1,\dots,
N,\; i\neq j.
\]
\item The vectors $\bG_i(z,\bt)$  satisfy
\begin{gather*}
\dfrac{\D\bG_i}{\D u_j}-\beta_{ij}\bG_j=0,\quad i,j=1,\dots,
N,\; i\neq j,\\
\frac{\D\bG_i}{\D u_i}+\sum_{\substack{k=1,\dots,N\\
k\neq i}}
 \beta_{ki}\bG_k=0, \quad i=1,\dots,N.
\end{gather*}
\item The coefficients $\beta_{ij}$ are solutions of the Lam\'{e}
equations:
\begin{align*}
&\frac{\partial\beta_{ij}}{\D u_k}-\beta_{ik}\beta_{kj}=0,\;\;
i,j,k=1,\dotsc, N,\;
\text{with $i,j,k$ different},
\\
&\frac{\partial\beta_{ij}}{\D
u_i}+
\frac{\partial\beta_{ji}}{\D u_j}+
\sum_{\substack{k=1,\dotsc,N\\ k\neq i,j}}
\beta_{ki}\beta_{kj}=0,\quad i,j=1,\dotsc,N,\;
i\neq j.
\end{align*}
\end{enumerate}
\end{Th}
\begin{proof}
\begin{enumerate}
\item
Form \eqref{bilbaker} we easily deduce that
\[
P_j\Big(\frac{\D\psi}{\D u_i}-\frac{\D G}{\D u_i}G^{-1}\psi\Big)\in
W,\quad i\neq j
\]
Then, because
\[
\dfrac{\D\psi}{\D u_i}-\dfrac{\D G}{\D
u_i}G^{-1}\psi\sim (P_i z+{\cal O}(1))\psi_0(z,\bt)
\]
from  statement 2 of Prop. \ref{asymptotic} we conclude the
identity
\[
P_j\dfrac{\D\psi}{\D u_i}=P_j\dfrac{\D G}{\D u_i}G^{-1}\psi,\quad
i,j=1,\dots,N,\; i\neq j.
\]
On the other hand:
\[
\int_{S^1}\frac{\dif z}{2\pi\I
z}\dfrac{\D\psi}{\D u_i}(z,\bt)\psi^\t(-z,\bt)=\dfrac{\D G}{\D
u_i}(\bt)G^{-1}(\bt)
\]
and evaluating the residue of the integrand at $z=\infty$ we get
\[
\beta_{ji}(\bt)E_{ji}=P_j\dfrac{\D G}{\D
u_i}(\bt)G^{-1}(\bt), \quad i,j=1,\dots,N,\; i\neq j,
\]
which gives the stated result.
\item Since $G(\bt)=\psi(0,\bt)$,
its rows  satisfy the same linear system as the rows of the Baker
function (see 1 of the Theorem). Moreover, as it is an orthogonal
matrix its rows  form an orthonormal frame $\{\bG_i\}_{i=1}^N$.
Hence, we have:
\[
\frac{\D\bG_i}{\D u_i}=\sum_{k=1}^N A_{ik}\bG_k
\]
with
\[
\frac{\D\bG_i}{\D u_i}\cdot\bG_k=A_{ik}.
\]
But from  $\bG_i\cdot\bG_k=\delta_{ik}$ it follows that
\[
\frac{\D\bG_k}{\D u_i}\cdot\bG_i=-A_{ik},\quad A_{ii}=0;
\]
and recalling that $\dfrac{\D\bG_k}{\D u_i}=\beta_{ki}\bG_i$ we get
\[
\frac{\D\bG_i}{\D u_i}+\sum_{\substack{k=1,\dots,N\\k\neq i}}
 \beta_{ki}\bG_k=0.
\]
\item The Lam\'{e} equations are the compatibility conditions for the
linear system satisfied by the rows of $G$.
\end{enumerate}
\end{proof}
\section{The Ribaucour transformation as a vertex operator}

In this section we are going to identify the action of the vertex
operator $\V_i(z)$ with the classical Ribaucour transformation. For
this aim we shall use some  identities for the Baker functions
which in turn correspond to Fay identities for the underlying
$\tau$-functions. If we set $\bt\mapsto \bt+[1/p]\be_i$
 and $\bt'\mapsto\bt$ in \eqref{bilbaker}, then taking into account that
 \[
\V_i(p)\psi(z,\bt)=\chi(z,\bt+[1/p]\be_i)\Big[1-\frac{2z}{z-p}P_i\Big]\psi_0(z,\bt),
\]
for $|p|>1$, we obtain
\begin{equation}\label{residuo}
 1-2P_i-\res_p(\V_i(p)\psi(z,\bt))P_i\psi^\t(-p,\bt)=(\V_i(p)G(\bt)) G(\bt)^\t.
 \end{equation}
 Now, the orthogonal character of the right-hand side of this
formula implies
\begin{multline*}
(1-2P_i)\psi(-p,\bt))P_i\res_p(\V_i(p)\psi^\t(z,\bt)) +
\res_p(\V_i(p)\psi(z,\bt))P_i \psi^\t(-p,\bt)(1-2P_i)\\=
\Big[\sum_{k=1}^N\psi_{ki}^2(-p,\bt)\Big]
 \res_p(\V_i(p)\psi(z,\bt))P_i
\res_p(\V_i(p)\psi^\t(z,\bt)),
\end{multline*}
By multiplying this expression on the left by $P_j$ and on the
right by $P_i$ it follows that
\begin{multline*}
(-1)^{\delta_{ij}}\psi_{ji}(-p,\bt))\res_p(\V_i(p)\psi_{ii}(z,\bt))
-\res_p(\V_i(p)\psi_{ji}(z,\bt)) \psi_{ii}(-p,\bt)\\=
\Big[\sum_{k=1}^N\psi_{ki}^2(-p,\bt)\Big] \res_p(\V_i(p)\psi_{ii}(z,\bt))
 \res_p(\V_i(p)\psi_{ji}(z,\bt)),
\end{multline*}
or equivalently
\begin{equation}\label{residuo2}
\res_p(\V_i(p)\psi_{ji}(z,\bt))=(-1)^{\delta_{ij}}
\frac{2 \psi_{ji}(-p,\bt)}{\sum_{k=1}^N\psi_{ki}^2(-p,\bt)}.
\end{equation}
If we multiply \eqref{residuo} by $P_j$ and $G(\bt)$ on the left
and right, respectively, we get
\[
(-1)^{\delta_{ij}}\bG_j(\bt)-
\res_p(\V_i(p)\psi_{ji}(z,\bt))\sum_{k=1}^N\psi_{ki}(-p,\bt)\bG_k(\bt)=
\V_i(p)\bG_j(\bt).
\]
This equation together with \eqref{residuo2} shows that
\[
(-1)^{\delta_{ij}}\V_i(p)\bG_j(\bt)=
\bG_j(\bt)-
\frac{2 \psi_{ji}(-p,\bt)}{\sum_{k=1}^N\psi_{ki}^2(-p,\bt)
}\Big[\sum_{k=1}^N\psi_{ki}(-p,\bt)\bG_k(\bt)\Big].
\]
Hence, we conclude
\begin{Th}
The Ribaucour transformation of the Lam\'{e} equations with
transformation data $\zeta_k(\bt):=\psi_{ki}(-p,\bt)$,
$k=1,\dots,N$ is given by
\[
{\cal R}(\bG_j)=(-1)^{\delta_{ij}}\V_i(p)\bG_j.
\]
\end{Th}

\end{document}